\documentclass{IET-Conf-Paper}

\usepackage{amsmath,amsfonts}
\usepackage{algorithmic}
\usepackage{algorithm}
\usepackage{array}
\usepackage{textcomp}
\usepackage{stfloats}
\usepackage{url}
\usepackage{verbatim}
\usepackage{graphicx}
\usepackage{csquotes}
\usepackage{amsmath,amssymb,amsfonts}
\usepackage{xcolor}
\usepackage{booktabs,tabularx}
\usepackage{array}
\usepackage{lineno}
\usepackage{hyperref}
\usepackage{makecell}
\usepackage{subcaption}
\usepackage{caption}
\usepackage{adjustbox}
\usepackage{bm}
\usepackage{multirow}
\usepackage{tikz}
\usepackage{colortbl}
\usepackage[table]{xcolor}

\header{25\textsuperscript{th} Wind \& Solar Integration Workshop | Porto, Portugal | 29 September -- 02 October 2026}

\makeatletter
\long\def\@tablecaption#1#2{\par%
  \setbox\tablecaptionbox=\vbox{\hsize\linewidth%
    {{#1}\enskip {#2}}\strut}%
    \unvbox\tablecaptionbox\par%
  \vspace{\belowcaptionskip}%
  }%
\makeatother

\begin{document}

\title{Quantifying Grid-Forming Requirements for System Strength: How Much and Where?}


\author{Ambuj Gupta\ad{1}\corr, Balarko Chaudhuri\ad{1}, Mark O'Malley\ad{1}}

\address{\add{1}{Department of Electrical and Electronic Engineering, Imperial College London, United Kingdom}
\add{*}{a.gupta23@imperial.ac.uk}}


\begin{abstract}
As power systems transition toward inverter-based resource (IBR)-dominated grids, Grid-ForMing (GFM) IBRs are increasingly expected to provide the voltage-source characteristics required to maintain adequate system strength. However, how many GFM IBRs are required, and where they should be located in the grid, remains unclear. This paper investigates the impact of the number and location of GFM IBRs on small-signal system strength. The Jacobian Voltage Stiffness Metric (JVSM), which quantifies the voltage-magnitude and voltage-phase stiffness from frequency-domain Jacobian characteristics, is used as a system-strength measure. The analysis demonstrates that increasing the number of GFM IBRs generally improves system strength; however, configurations with the same number of GFM IBRs can exhibit significantly different bus-wise system-strength characteristics depending on their locations. Thus, this paper provides a quantitative basis for specifying GFM requirements in terms of the resulting system strength, rather than solely in terms of penetration level or the number of GFM IBRs. Case studies on a modified IEEE 39-bus system demonstrate how JVSM can be used to quantify and compare the system-strength impact of different numbers and locations of GFM IBRs, with the resulting strength assessments validated against small-signal oscillatory stability.
\end{abstract}

\maketitle
\section{Introduction}

The transition toward inverter-based resource (IBR)-dominated power systems is reducing the synchronous generation that has traditionally provided voltage-magnitude and phase stiffness. In the sub-transient to transient time frame, synchronous generators (SGs) inherently exhibit voltage-source-behind-an-impedance (VSBI) characteristics. As SGs are displaced by IBRs, grid-forming (GFM) IBRs are expected to provide similar voltage-source characteristics and thereby support system strength \cite{aemo2023voluntary, ERCOT2024AGS_ESR, miso2024gridforming, ESIG2025_GFM, unifi2026specs}.

System strength describes the ability of the power system to maintain and control the voltage waveform following disturbances \cite{aemo2020systemstrength,nerc2017ibrlowshortcircuitstrength}. Traditionally, it has been quantified using short-circuit-level-based measures such as the short-circuit ratio (SCR). However, in IBR-dominated systems, system strength depends not only on network impedance and available fault level, but also on the dynamic characteristics of IBRs and their controls \cite{nerc2017ibrlowshortcircuitstrength, dozein2025systemstrength, huang2026systemstrength}. Thus, conventional SCR-based measures do not necessarily characterize small-signal voltage stiffness in IBR-dominated grids.

Because GFM IBRs are expected to provide the voltage-source characteristics needed to support system strength, this raises the practical question of how many GFM IBRs are required and where they should be located. Requirements based solely on installed capacity, number of GFM IBRs, or GFM penetration do not directly characterize the resulting system strength. The VSBI characteristics of a GFM IBR depend on its control design, while its system-level contribution also depends on its electrical location. Hence, this paper provides a quantitative basis for specifying GFM requirements that account for both the amount and location of the capability provided.

Several studies have investigated GFM requirements or placement from the perspective of small-signal stability and system strength. Generalized-short-circuit-ratio-based approaches can assess the number of GFM converters required for a specified stability margin \cite{10508461}, while impedance- and admittance-based approaches can identify weak-grid locations or devices contributing to poorly damped modes \cite{zhu2024impedance, hadjileonidas2024admittance}. However, these approaches either assume ideal voltage-source characteristics or require detailed system models, vector fitting, eigenvalue analysis, or reference impedance models, and the combined impact of GFM number and location on the resulting bus-wise voltage stiffness remains insufficiently analyzed.

In \cite{gupta2026JVSM}, the Jacobian Voltage Stiffness Metric (JVSM) was proposed to quantify small-signal VSBI strength at both the device and grid levels. At the grid level, JVSM provides a bus-wise measure of voltage stiffness and was shown to identify weak-grid regions vulnerable to poorly damped converter-driven oscillations. Building on this metric, this paper systematically investigates how the number and location of GFM IBRs affect small-signal system strength. The main contributions are:

\begin{itemize}

    \item JVSM-based strength assessments are validated against small-signal modal analysis across multiple GFM IBR distributions in the system.

    \item The impact of the number of GFM units on system strength is quantified by varying the number of GFM IBRs and evaluating the resulting bus-wise JVSM.

    \item The impact of GFM IBR location is assessed for the same number of GFM IBRs, showing that different placements can produce significantly different system-strength characteristics.

\end{itemize}

\section{VSBI Characterization of System Strength}

Voltage-source-behind-an-impedance (VSBI) characteristics describe the ability of a device, or of the grid as seen from a bus, to maintain its voltage magnitude and phase around an operating point following disturbances. For an ideal VSBI, the internal voltage remains constant in the sub-transient to transient time frame, while terminal-voltage disturbances are opposed through the corresponding exchange of active and reactive power. Thus, a rapid and significant power response indicates strong voltage stiffness, whereas a slower or lower-magnitude response indicates weaker VSBI characteristics \cite{gupta2026JVSM}.

\begin{figure}[htbp]
    \centering
    \includegraphics[width=\columnwidth]{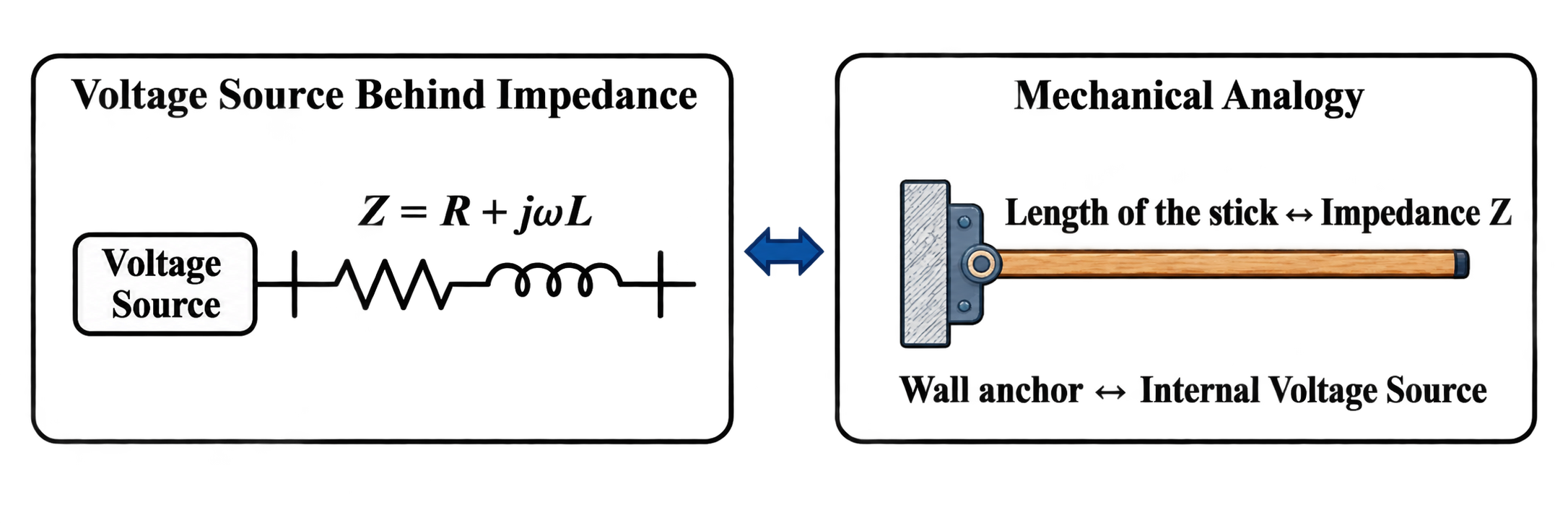}
    \caption{Mechanical analogy for ideal and non-ideal VSBI characteristics. The wall anchor represents the internal voltage source and the stick length represents the impedance $Z$.}
    \label{fig:vsbi_analogy}
\end{figure}

An intuitive interpretation of ideal and non-ideal VSBI characteristics can be obtained using the wall-mounted-stick analogy illustrated in Fig.~\ref{fig:vsbi_analogy}. The wall anchor represents the internal voltage source, while the length of the stick represents the impedance $Z$ between the internal source and the terminal. For an ideal voltage source, the wall anchor is perfectly rigid and remains fixed following a disturbance at the free end of the stick. The resulting stiffness at the free end is therefore determined by the stick length: a shorter stick, corresponding to a lower impedance, exhibits smaller oscillations following a disturbance, whereas a longer stick exhibits larger oscillations. This reflects the traditional understanding of system strength in synchronous-generator-dominated grids, where the internal voltage source is assumed to remain constant over the sub-transient time frame, and strength can therefore be characterized primarily through the impedance, as in SCR-based measures. In IBR-dominated grids, however, the voltage-source characteristics of a GFM IBR, or of the grid as observed from a bus, can be non-ideal because the underlying voltage-source response is control-dependent. In the analogy, this corresponds to a weak or loose wall anchor. Consequently, even for the same stick length, the oscillations at the free end can be substantially different depending on the stiffness of the wall anchor. Similarly, identical network impedances can result in different voltage stiffness depending on the dynamic characteristics of the underlying voltage source, motivating a strength characterization that captures both the impedance and the non-ideal voltage-source response.

For small disturbances, the voltage-magnitude, $V$, and phase-angle, $\theta$, perturbations are related to the active- and reactive-power responses through the frequency-domain Jacobian given in \eqref{eq:jacobian}.

\begin{equation}
\begin{bmatrix}
P(s)\\
Q(s)
\end{bmatrix}
=
\underbrace{
\begin{bmatrix}
J_{PV}(s) & J_{P\theta}(s)\\
J_{QV}(s) & J_{Q\theta}(s)
\end{bmatrix}}_{\bm{J}(s)}
\begin{bmatrix}
V(s)\\
\theta(s)
\end{bmatrix}.
\label{eq:jacobian}
\end{equation}

where the elements of $\bm{J}(s)$ represent the corresponding small-signal frequency-domain transfer characteristics. In particular,
$J_{P\theta}(s)=\left.P(s)/\theta(s)\right|_{V(s)=0}$ represents the active-power response to a voltage-phase perturbation, while
$J_{QV}(s)=\left.Q(s)/V(s)\right|_{\theta(s)=0}$ represents the reactive-power response to a voltage-magnitude perturbation.

In bulk transmission systems with high $X/R$ ratios, the dominant couplings are between active power and voltage phase, and between reactive power and voltage magnitude. Voltage-phase and voltage-magnitude stiffness are therefore characterized by $J_{P\theta}(s)$ and $J_{QV}(s)$, respectively; $J_{PV}(s)$ and $J_{Q\theta}(s)$ are not considered here.

\section{Jacobian Voltage Stiffness Metric}
\subsection{Metric Definition}

To quantify the voltage stiffness described in Section~II, the Jacobian Voltage Stiffness Metric (JVSM) proposed in \cite{gupta2026JVSM} combines the magnitude and speed of the active- and reactive-power responses to voltage disturbances. The voltage-magnitude stiffness, $\mathrm{JVSM}_{V}$, is calculated
from the $J_{QV}(s)$ characteristic as

\begin{equation}
\mathrm{JVSM}_{V}
=
\frac{
\displaystyle \int_{f_{\min}}^{f_{\max}}
\left|J_{QV}(j2\pi f)\right|
\log_{10}(f)\,df
}{
\displaystyle \int_{f_{\min}}^{f_{\max}}
\log_{10}(f)\,df
}.
\label{eq:jvsm_v}
\end{equation}

The voltage-phase stiffness, $\mathrm{JVSM}_{\theta}$, is similarly obtained from the $J_{P\theta}(s)$ characteristic as

\begin{equation}
\mathrm{JVSM}_{\theta}
=
\frac{
\displaystyle \int_{f_{\min}}^{f_{\max}}
\left|J_{P\theta}(j2\pi f)\right|
\log_{10}(f)\,df
}{
\displaystyle \int_{f_{\min}}^{f_{\max}}
\log_{10}(f)\,df
}.
\label{eq:jvsm_theta}
\end{equation}

The two components can be combined into a single voltage-stiffness metric,

\begin{equation}
\mathrm{JVSM}
=
\sqrt{
\frac{
\mathrm{JVSM}_{\theta}^{2}
+
\mathrm{JVSM}_{V}^{2}
}{2}
}.
\label{eq:jvsm}
\end{equation}

The Jacobian magnitudes quantify the strength of the power response to voltage disturbances, while the $\log_{10}(f)$ weighting emphasizes
higher-frequency, and therefore faster, responses. The magnitudes are evaluated in p.u. rather than in dB, such that large responses are retained directly in the metric. Since VSBI characteristics are associated with the
sub-transient to transient time frame, approximately $0$--$200$~ms, the frequency range is selected as $f_{\min}=5$~Hz and $f_{\max}=100$~Hz. The lower bound excludes slower dynamics, while the upper bound limits the contribution of faster dynamics outside the VSBI range of interest.

Larger $\mathrm{JVSM}_{V}$ and $\mathrm{JVSM}_{\theta}$ therefore indicate
stronger voltage-magnitude and voltage-phase stiffness, respectively, while
the combined JVSM provides a single measure of small-signal VSBI strength.

A key advantage of JVSM is that it can be calculated directly from frequency-domain scans of a device or grid without requiring detailed internal models, a reference impedance, or prior classification of an IBR as GFM or GFL. Its calculation also avoids vector fitting and the subsequent eigenvalue- or singular-value-based analysis required by several existing dynamic strength metrics. In addition, JVSM retains the physically relevant $J_{P\theta}(s)$ and $J_{QV}(s)$ characteristics rather than aggregating the full admittance matrix into a single intermediate quantity, thereby preserving the voltage-phase and voltage-magnitude coupling dynamics that dominate in high-$X/R$ transmission systems. These features make JVSM particularly suitable as a computationally simple pre-screening metric for comparing the small-signal voltage stiffness of different devices, operating conditions, and grid configurations.

\subsection{Device- and Grid-Level Interpretation of JVSM}

JVSM can be applied at two levels: to quantify the standalone voltage stiffness of an individual IBR and the resulting grid strength observed from a particular bus. This distinction is important because the device-level VSBI strength of a GFM IBR does not directly determine the system strength throughout the network.

At the device level, JVSM is calculated from the $J_{P\theta}(s)$ and $J_{QV}(s)$ characteristics obtained from frequency-domain scans at the terminals of a standalone IBR. The resulting $\mathrm{JVSM}_{\theta}$, $\mathrm{JVSM}_{V}$, and combined JVSM quantify its voltage-phase, voltage-magnitude, and overall small-signal VSBI strength, respectively. These characteristics depend on the IBR dynamics and control design, allowing the VSBI capability of different IBRs to be compared independently of their GFM or GFL classification.

At the grid level, JVSM is calculated from the driving-point characteristics looking into the network from bus $i$. For a given operating point, the $2\times2$ frequency-dependent driving-point admittance, $\bm{Y}^{\mathrm{Grid}}_{i}(s)$, can be obtained directly from a whole-system frequency scan at bus $i$. Alternatively, when the network topology, parameters, and frequency-dependent admittance characteristics of all connected devices are available, the grid driving-point admittance can be calculated without performing a whole-system frequency scan. At each frequency, the network admittances are combined with the terminal admittance characteristics of the connected devices to form the frequency-dependent system admittance representation \cite{gabo_2026}. The driving-point admittance at bus $i$ is then obtained from this whole-system representation, thereby capturing the combined effect of the network and all connected device dynamics as observed from that bus. Repeating this calculation over the frequency range of interest and for each bus provides the bus-wise driving-point admittance characteristics required for the JVSM calculation.

The resulting $\bm{Y}^{\mathrm{Grid}}_{i}(s)$ is converted to the corresponding driving-point Jacobian using the operating-point quantities at bus $i$, from which $\mathrm{JVSM}^{\mathrm{Grid}}_{i}$ is calculated using \eqref{eq:jvsm_v}--\eqref{eq:jvsm}. Accordingly, the standalone strength of IBR $k$ is denoted by $\mathrm{JVSM}^{\mathrm{IBR}}_{k}$, while the system strength observed from bus $i$ is denoted by $\mathrm{JVSM}^{\mathrm{Grid}}_{i}$.

\section{Case Studies: GFM Requirements for System Strength}
\subsection{Test System}

\begin{figure}[htbp]
    \centering
    \includegraphics[width=0.9\columnwidth]{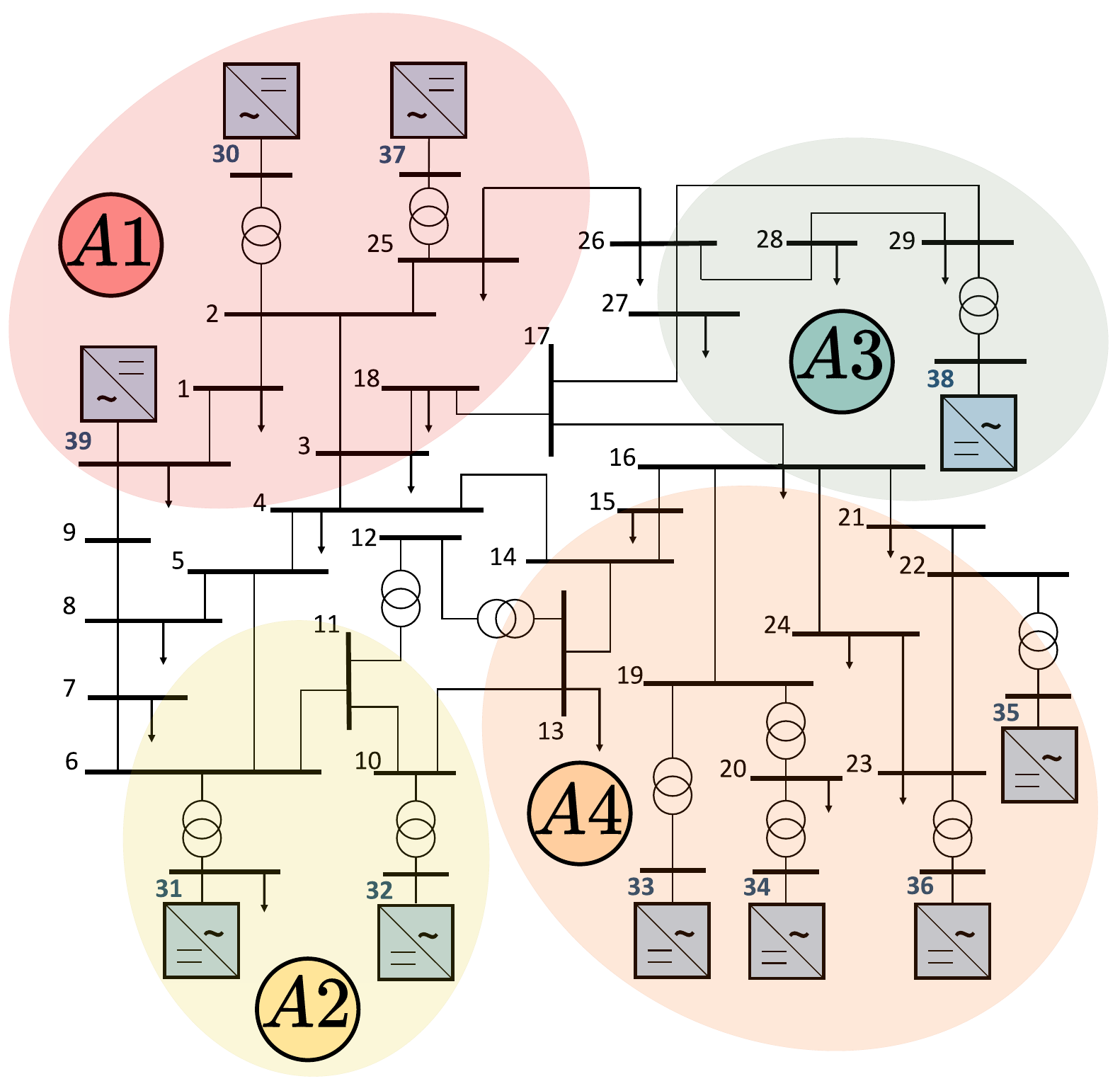}
    \caption{Modified IEEE 39-bus system with all synchronous generators replaced by IBRs at buses 30--39. The IBR connection points are grouped into four broad areas, $A_1$--$A_4$.}
    \label{fig:ieee39}
\end{figure}

The impact of GFM number and location on system strength is investigated using a modified IEEE 39-bus system. As shown in Fig.~\ref{fig:ieee39}, all ten synchronous generators in the original system are replaced by IBRs connected at buses 30--39, resulting in a 100\% IBR-based system. Across all configurations, IBR ratings are fixed; units are switched between GFL and GFM, so only GFM number and location vary.

For the locational analysis, the ten IBR locations are roughly grouped into four broad areas based on their electrical proximity within the network, as illustrated in Fig.~\ref{fig:ieee39}. Area~1 comprises buses 30, 37, and 39; Area~2 comprises buses 31 and 32; Area~3 comprises bus 38; and Area~4 comprises buses 33--36. These areas are used subsequently to interpret the locational distribution of GFM IBRs and its effect on system strength.

\subsection{JVSM and Modal Assessment}

First, four representative two-GFM configurations are considered to illustrate the JVSM-based system-strength assessment and its relation to small-signal oscillatory characteristics. In all cases, two of the ten IBRs operate as GFM units and the remaining eight as GFL units, with identical GFM control characteristics.

\begin{figure*}[!t]
    \centering
    \includegraphics[width=\textwidth]
    {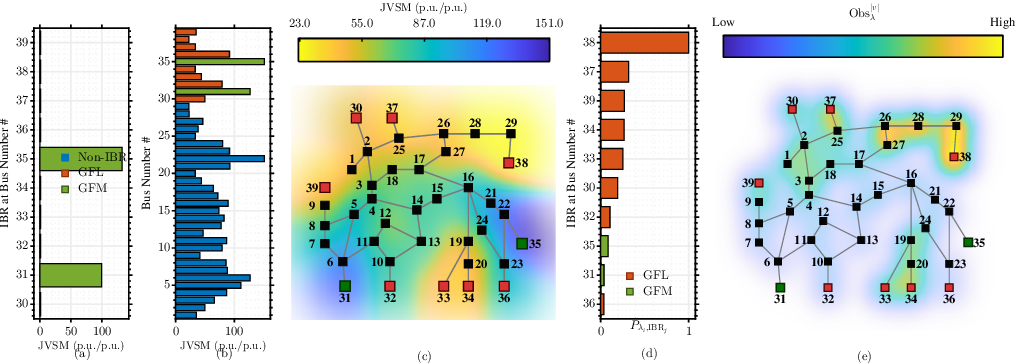}
    \caption{JVSM-based strength assessment and modal validation for the representative two-GFM configuration with GFM IBRs at buses 31 and 35. (a) Standalone JVSM of the IBRs at buses 30--39. (b) Bus-wise grid JVSM, with non-IBR, GFL-IBR, and GFM-IBR buses distinguished. (c) Spatial distribution of bus-wise JVSM. (d) Normalized IBR participation factors for the critical oscillatory mode. (e) Voltage-magnitude modal observability of the same mode.}
    \label{fig:case_3135}
\end{figure*}

To provide balanced spatial coverage of the four areas in Fig.~\ref{fig:ieee39}, representative IBR locations are selected at bus~39 in $A_1$, bus~31 in $A_2$, bus~38 in $A_3$, and bus~35 in $A_4$. The four cases use GFM pairs $(31,35)$, $(35,38)$, $(38,39)$, and $(31,39)$, such that each area contains a GFM IBR in exactly two cases. The cases are selected for balanced spatial coverage rather than their JVSM or damping values, and are used to demonstrate the physical interpretation and repeatability of the assessment. The $(31,35)$ configuration is used as the reference case in Fig.~\ref{fig:case_3135}. Fig.~\ref{fig:case_3135}(a)--(e) summarizes the device-level JVSM, bus-wise grid strength, and modal characteristics used to assess the corresponding small-signal stability.

Fig.~\ref{fig:case_3135}(a) shows the standalone JVSM of the IBR connected at each generator bus. These values quantify the voltage-stiffness capability provided by the individual devices. As expected, the GFM IBRs at buses 31 and 35 exhibit substantially larger $\mathrm{JVSM}^{\mathrm{IBR}}$ than the GFL IBRs, reflecting their stronger VSBI characteristics.

The resulting bus-wise grid strength is shown in Fig.~\ref{fig:case_3135}(b) and spatially in Fig.~\ref{fig:case_3135}(c). Unlike the standalone device-level values, $\mathrm{JVSM}^{\mathrm{Grid}}_i$ reflects the combined effect of the network and all connected IBRs. Higher JVSM values indicate stronger local voltage stiffness, with stronger regions observed around the GFM IBRs at buses 31 and 35 and weaker regions elsewhere in the network. The weakest GFL-connected location is bus~38, illustrating that the strength contribution of a GFM IBR is predominantly local rather than uniform across the system.

Since GFL IBRs are particularly sensitive to weak-grid conditions, the minimum GFL-bus JVSM is used as a scalar strength indicator, $\mathrm{JVSM}^{\mathrm{GFL}}_{\min}=\min_{i\in\mathcal{B}_{\mathrm{GFL}}}\mathrm{JVSM}^{\mathrm{Grid}}_{i}$, where $\mathcal{B}_{\mathrm{GFL}}$ is the set of GFL-connected buses. For the reference case, $\mathrm{JVSM}^{\mathrm{GFL}}_{\min}=22.43$ at bus~38. This conservative measure of the weakest GFL-bus voltage stiffness is used subsequently to compare system configurations.

The IBR contribution to the critical oscillatory mode is quantified from the left and right eigenvectors, $\bm{\phi}_i$ and $\bm{\psi}_i$, obtained from the white-box linearized small-signal model of the full system, by combining the element-wise participation products over the states of each IBR and normalizing across all IBRs to obtain $\widetilde{P}_{\lambda_i,\mathrm{IBR}_j}$. Fig.~\ref{fig:case_3135}(d) shows that the GFM IBRs have low participation, while the GFL IBRs located in the weak-grid areas have the highest contribution to the critical mode. In particular, the GFL IBR at bus~38 contributes most strongly and is also the GFL-connected bus with the minimum JVSM in Fig.~\ref{fig:case_3135}(b)--(c).

The spatial manifestation of the critical mode is assessed using voltage-magnitude modal observability, $\bm{\mathrm{Obs}}^{|V|}_{\lambda_i}=\bm{C}_{G}\bm{\psi}_i$, where $\bm{C}_{G}$ maps the system states to bus-voltage outputs. Fig.~\ref{fig:case_3135}(e) shows the corresponding relative spatial distribution, with the mode most strongly observable in the weaker network regions, particularly around bus~38. This independently corroborates both the participation-factor result in Fig.~\ref{fig:case_3135}(d) and the weak region identified by the JVSM map in Fig.~\ref{fig:case_3135}(c).

The same assessment is repeated for the other three selected two-GFM configurations, with the GFM-supported areas shuffled. As summarized in Table~\ref{tab:representative_2gfm}, a consistent spatial pattern is observed: for GFM buses $(31,39)$ in $A_2$--$A_1$, the weakest GFL bus is 38 in $A_3$; for $(35,38)$ in $A_4$--$A_3$, it shifts to bus~39 in $A_1$; for $(38,39)$ in $A_3$--$A_1$, it occurs at bus~34 in $A_4$; and for $(31,35)$ in $A_2$--$A_4$, at bus~38 in $A_3$.

\begin{table}[htbp]
\centering
\caption{Selected two-GFM illustrative cases.}
\label{tab:representative_2gfm}
\setlength{\tabcolsep}{3pt}
\resizebox{\columnwidth}{!}{%
\begin{tabular}{cccc}
\toprule
\makecell{GFM buses\\(Areas)} &
\makecell{Weakest GFL bus\\(Area, $\mathrm{JVSM}^{\mathrm{GFL}}_{\min}$)} &
$\zeta_{\mathrm{worst}}$ (\%) &
$f_{\mathrm{worst}}$ (Hz) \\
\midrule
31 ($A_2$), 39 ($A_1$) & 38 ($A_3$, 21.27) & -7.51 & 7.13 \\
35 ($A_4$), 38 ($A_3$) & 39 ($A_1$, 26.66) & 15.88 & 6.90 \\
38 ($A_3$), 39 ($A_1$) & 34 ($A_4$, 25.00) & -4.59 & 7.24 \\
31 ($A_2$), 35 ($A_4$) & 38 ($A_3$, 22.43) & 11.70 & 7.39 \\
\bottomrule
\end{tabular}%
}
\end{table}


Across all four cases, higher-JVSM regions follow GFM-supported areas, while the weakest GFL buses lie in areas without direct GFM provision. The high IBR participation and modal observability show that bus-wise JVSM can pre-screen GFL-connected weak-grid locations associated with poorly damped modes and identify the network regions where the resulting voltage oscillations manifest most strongly. The following subsections quantify the effects of GFM number and location across the complete configuration set.

\subsection{Impact of GFM Number}

Having established the interpretation of bus-wise JVSM for individual configurations, the impact of GFM number is assessed by varying the number of GFM IBRs from one to four while keeping their device-level characteristics unchanged. All possible combinations of the ten IBR connection points are considered, giving 10, 45, 120, and 210 cases for one, two, three, and four GFM IBRs, respectively, and 385 configurations in total. This allows the effect of a higher number of GFM IBRs to be assessed across all possible placements rather than for selected locations.

\begin{figure}[htbp]
    \centering
    \includegraphics[width=0.9\columnwidth]
    {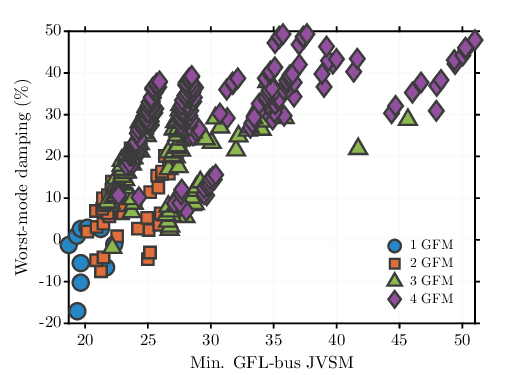}
    \caption{Worst-mode damping versus minimum GFL-bus JVSM for all configurations, grouped by the number of GFM IBRs.}
    \label{fig:gfm_number_scatter}
\end{figure}

Fig.~\ref{fig:gfm_number_scatter} compares the worst-mode damping ratio of each configuration with its $\mathrm{JVSM}^{\mathrm{GFL}}_{\min}$. A clear positive relationship is observed: configurations with low minimum GFL-bus JVSM generally exhibit low or negative damping, while configurations with larger $\mathrm{JVSM}^{\mathrm{GFL}}_{\min}$ exhibit more strongly damped critical modes. Increasing the number of GFM IBRs progressively shifts the configurations towards higher JVSM and higher damping. The one-GFM cases are concentrated towards the lower-left region, whereas the four-GFM cases extend furthest towards the upper-right region.

The relationship in Fig.~\ref{fig:gfm_number_scatter} can also be used to translate a desired small-signal stability margin into a minimum system-strength requirement. For example, for the configurations considered here, a target worst-mode damping ratio of at least $10\%$ could be associated with a conservative requirement of $\mathrm{JVSM}^{\mathrm{GFL}}_{\min}\geq 30$. Equivalently, a minimum JVSM of 30 could be required at each GFL-connected bus, and the number of GFM IBRs could then be selected to satisfy this system-level strength requirement. This provides an alternative to prescribing GFM requirements solely in terms of the number or total installed capacity of GFM IBRs. The specific JVSM threshold is system- and device-dependent, but the scatter relationship provides a systematic means of deriving such requirements for a specified stability margin.

\begin{figure}[htbp]
    \centering
    \includegraphics[width=0.9\columnwidth]
    {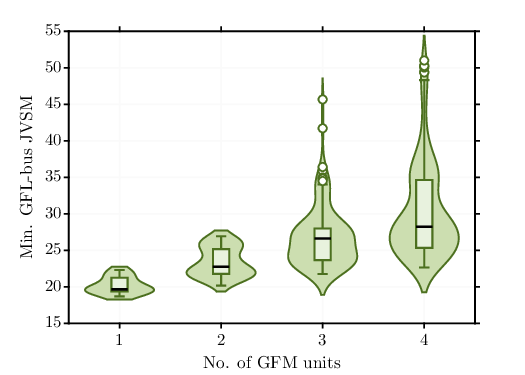}
    \caption{Distribution of the minimum GFL-bus JVSM for different numbers of GFM IBRs.}
    \label{fig:gfm_number_jvsm}
\end{figure}

The effect of GFM number on system strength is shown explicitly in Fig.~\ref{fig:gfm_number_jvsm}. The distribution of $\mathrm{JVSM}^{\mathrm{GFL}}_{\min}$ shifts upward as additional GFM IBRs are introduced, with the median increasing from 19.66 for one GFM IBR to 22.94, 26.63, and 28.24 for two, three, and four GFM IBRs, respectively. Increasing the number of GFM IBRs therefore generally increases the voltage stiffness. However, the number of GFM IBRs alone does not uniquely determine the resulting system strength.

\begin{figure}[htbp]
    \centering
    \includegraphics[width=0.9\columnwidth]
    {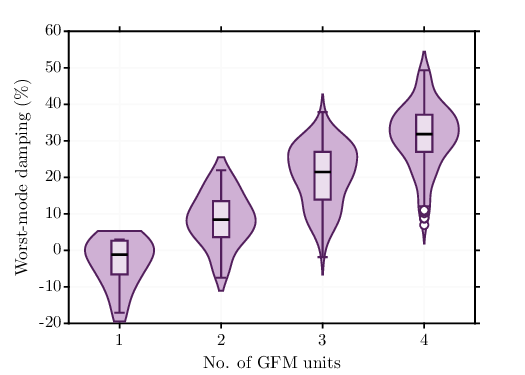}
    \caption{Distribution of the worst-mode damping ratio for different numbers of GFM IBRs.}
    \label{fig:gfm_number_damping}
\end{figure}

A corresponding improvement in small-signal stability is observed in Fig.~\ref{fig:gfm_number_damping}. The median worst-mode damping increases from $-1.15\%$ for the one-GFM cases to $8.52\%$, $21.46\%$, and $31.83\%$ for the two-, three-, and four-GFM cases, respectively. Six of the ten one-GFM configurations contain a negatively damped mode, compared with five of the 45 two-GFM configurations and one of the 120 three-GFM configurations. For the GFM characteristics considered here, all 210 four-GFM configurations are positively damped.

Overall, increasing the number of GFM IBRs generally improves the minimum GFL-bus voltage stiffness and damping, but GFM number alone does not determine system strength. The next subsection therefore examines location at a fixed GFM number.

\subsection{Impact of GFM Location}

\begin{figure*}[htbp]
    \centering

    \begin{subfigure}[t]{\textwidth}
        \centering
        \includegraphics[width=\linewidth]
        {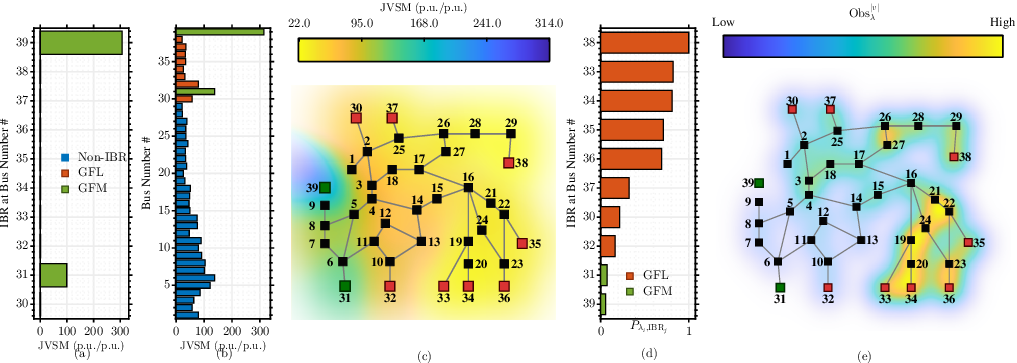}
        \caption{Base configuration with GFM IBRs at buses 31 and 39.}
        \label{fig:loc_2gfm}
    \end{subfigure}
    \hfill
    \begin{subfigure}[t]{\textwidth}
        \centering
        \includegraphics[width=\linewidth]
        {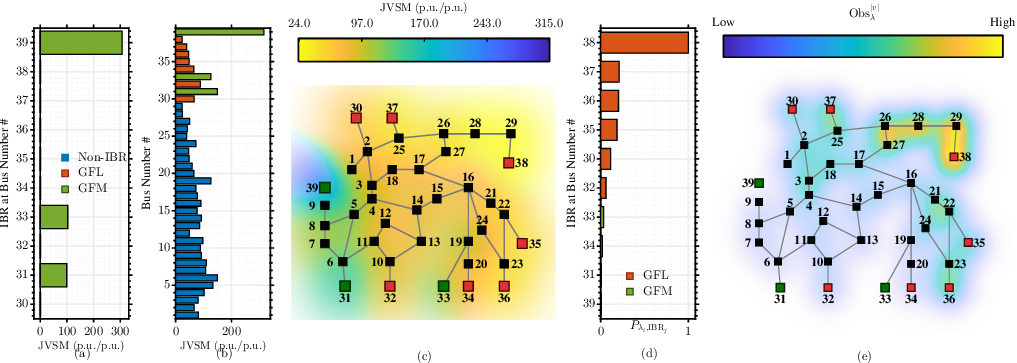}
        \caption{GFM IBRs at buses 31, 33, and 39.}
        \label{fig:loc_3gfm}
    \end{subfigure}
    \hfill
    \begin{subfigure}[t]{\textwidth}
        \centering
        \includegraphics[width=\linewidth]
        {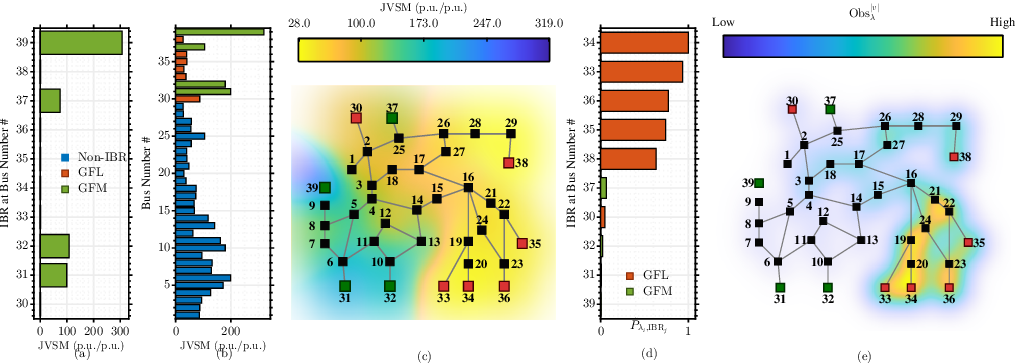}
        \caption{GFM IBRs at buses 31, 32, 37, and 39.}
        \label{fig:loc_4gfm}
    \end{subfigure}

    \caption{JVSM-based strength and modal assessment for the three illustrative GFM-location configurations. In each case, IBR participation factors and voltage-magnitude modal observability are normalized by their respective maxima within that configuration and therefore indicate relative values within the case, rather than absolute magnitudes for comparison across configurations.}
\end{figure*}

The effect of GFM location is first illustrated using the worst-damped two-GFM configuration as a base case. The GFM IBRs are connected at buses 31 and 39, in $A_2$ and $A_1$, respectively, while the remaining eight IBRs operate as GFL units. The weakest GFL-connected location is bus~38 in $A_3$, with
$\mathrm{JVSM}^{\mathrm{GFL}}_{\min}=21.27$. The corresponding worst-damped mode occurs at $7.13$~Hz with a damping ratio of $-7.51\%$, and the system is therefore small-signal unstable. The JVSM and modal characteristics of this base configuration are shown in Fig.~\ref{fig:loc_2gfm}.

Starting from this configuration, two alternatives are considered for increasing the number of GFM IBRs. In the first, one additional GFM IBR is introduced at bus~33, giving GFM buses $(31,33,39)$. Importantly, bus~33 lies in $A_4$, which was not supported by a GFM IBR in the base case. The three GFM IBRs are therefore distributed across three of the four network areas, $A_1$, $A_2$, and $A_4$. The minimum GFL-bus JVSM increases from 21.27 to 23.47, while the worst-mode damping improves substantially from $-7.51\%$ to $19.00\%$. The critical-mode frequency also shifts from $7.13$~Hz to $7.92$~Hz. The corresponding JVSM and modal assessment is shown in Fig.~\ref{fig:loc_3gfm}.

A contrasting case is obtained by adding two GFM IBRs to the same base configuration, at buses 32 and 37. This results in four GFM IBRs at buses $(31,32,37,39)$. However, buses 31 and 32 both belong to $A_2$, while buses 37 and 39 both belong to $A_1$. Consequently, although the number of GFM IBRs is doubled relative to the base case, the GFM IBRs remain concentrated within the same two network areas. For this configuration, $\mathrm{JVSM}^{\mathrm{GFL}}_{\min}$ increases to 27.10 and the system becomes stable, with a worst-mode damping ratio of $8.73\%$ at $7.89$~Hz, as shown in Fig.~\ref{fig:loc_4gfm}.

Despite containing one fewer GFM IBR, the $(31,33,39)$ configuration achieves higher worst-mode damping ($19.00\%$) than $(31,32,37,39)$ ($8.73\%$). Bus~33 extends GFM support into previously unsupported $A_4$, whereas buses~32 and 37 reinforce already-supported $A_2$ and $A_1$. Thus, the benefit of adding GFM IBRs depends on both their number and location.

To assess this locational effect systematically while holding the GFM number fixed, the remainder of this subsection considers all 210 four-GFM configurations. A simple GFM spread score, $S_{\mathrm{GFM}}$, is introduced as the number of the four network areas containing at least one GFM IBR. Therefore, for the four-GFM cases, $S_{\mathrm{GFM}}$ ranges from one, where all GFM IBRs are concentrated within a single area, to four, where at least one GFM IBR is present in every network area. The spread score is used only as a simple indicator of spatial distribution; configurations having the same score can still contain substantially different electrical placements.

\begin{figure}[htbp]
    \centering
    \includegraphics[width=0.9\columnwidth]
    {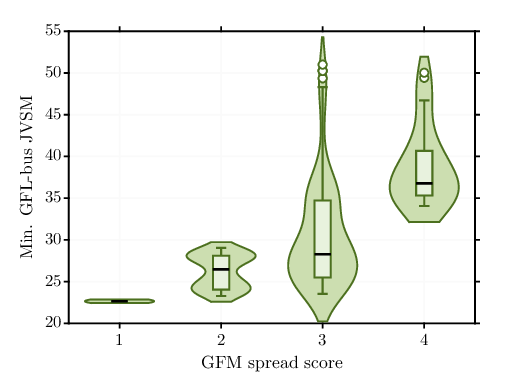}
    \caption{Distribution of the minimum GFL-bus JVSM for the
    four-GFM configurations grouped by GFM spread score.}
    \label{fig:gfm_location_jvsm}
\end{figure}

Fig.~\ref{fig:gfm_location_jvsm} shows that increasing the spatial spread of the four GFM IBRs generally increases the minimum voltage
stiffness experienced by the GFL IBRs. The median $\mathrm{JVSM}^{\mathrm{GFL}}_{\min}$ increases from 22.66 for $S_{\mathrm{GFM}}=1$ to 26.48, 28.29, and 36.77 for spread scores
of two, three, and four, respectively. In particular, configurations in which the four GFM IBRs are distributed across all four areas exhibit a substantial upward shift in the minimum GFL-bus strength. The spread within each group also shows that the number of areas containing GFM IBRs is only a coarse measure of location; different placements within the same spread score can still produce different system-strength outcomes.

\begin{figure}[htbp]
    \centering
    \includegraphics[width=0.9\columnwidth]
    {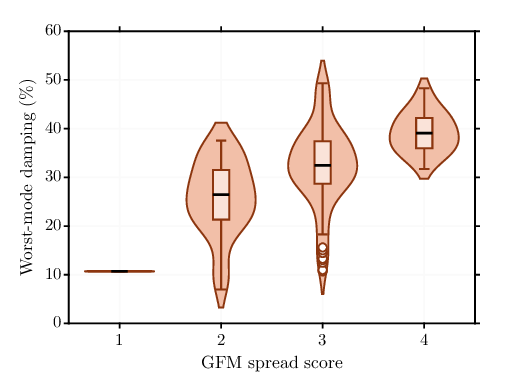}
    \caption{Distribution of the worst-mode damping ratio for the
    four-GFM configurations grouped by GFM spread score.}
    \label{fig:gfm_location_damping}
\end{figure}

A similar trend is observed in the corresponding small-signal
stability results in Fig.~\ref{fig:gfm_location_damping}. The median
worst-mode damping ratios increase from $10.71\%$ to $26.45\%$,
$32.46\%$, and $39.09\%$ as the spread score increases from one to four. Thus, even though every configuration contains exactly four
GFM IBRs with identical device-level characteristics, substantially different stability characteristics result depending on how the four GFM IBRs are distributed throughout the network. As with the JVSM results, the variation within individual spread-score groups shows that the particular GFM buses remain important in addition to the overall spatial spread.

These results show that the number and spatial distribution of GFM IBRs must be considered jointly when specifying system-strength requirements. A system-level requirement based on the resulting bus-wise voltage stiffness, rather than GFM number alone, more directly assesses whether the selected GFM configuration provides sufficient and appropriately distributed system strength.


\section{Conclusion}

This paper uses JVSM to quantify how both the number and location of GFM IBRs shape small-signal system strength in an IBR-dominated grid. System strength cannot be determined from the number of GFM IBRs alone. Configurations with the same number of GFM IBRs can produce substantially different voltage-stiffness and small-signal stability characteristics depending on their locations, while more spatially distributed GFM IBRs generally strengthen the weakest GFL-connected regions and improve damping. GFM requirements should therefore be expressed in terms of the resulting system strength, rather than solely through penetration level, installed capacity, or number of GFM IBRs. Bus-wise JVSM provides a practical basis for such requirements. It can be used to screen the grid for weak locations, compare candidate GFM placements, and assess whether a specified minimum level of voltage stiffness is achieved across GFL-connected buses. It also identifies the network regions most vulnerable to poorly damped oscillations, providing a computationally simple pre-screening step before detailed stability analysis. This enables the GFM requirement to be treated as a system-strength problem: how many GFM IBRs are required, and where should they be located? Future work can extend the framework to jointly consider GFM number, capacity, location, device characteristics, and operating conditions, providing a more comprehensive basis for system-strength planning.

\section{Acknowledgements}
This work was funded by a Leverhulme International Professorship, grant reference [LIP-2020-002] and by the Engineering and Physical Sciences Research Council [EP/Y025946/1]

\section{References}
\bibliographystyle{IEEEtran}
\bibliography{JVSM_References}

\end{document}